\documentclass[a4paper]{article}

\usepackage{INTERSPEECH2020}
\usepackage{cite,graphicx}
\title{Stacked 1D convolutional networks \\for end-to-end small footprint voice trigger detection}
\name{Takuya Higuchi, Mohammad Ghasemzadeh, Kisun You, Chandra Dhir}
\address{
  Apple}
\email{$\lbrace$takuya$\_$higuchi, mghasemzadeh, kisun$\_$you, cdhir$\rbrace$@apple.com}

\begin{document}

\maketitle
\begin{abstract}
  We propose a  stacked 1D convolutional neural network (S1DCNN) for end-to-end small footprint voice trigger detection in a streaming scenario. Voice trigger detection is an important speech application, with which users can activate their devices by simply saying a keyword or phrase. Due to privacy and latency reasons, a voice trigger detection system should run on an always-on processor on device. Therefore, having small memory and compute cost is crucial for a voice trigger detection system. Recently, singular value decomposition filters (SVDFs) has been used for end-to-end voice trigger detection. The SVDFs approximate a fully-connected layer with a low rank approximation, which reduces the number of model parameters. In this work, we propose S1DCNN as an alternative approach for end-to-end small-footprint voice trigger detection. An S1DCNN layer consists of a 1D convolution layer followed by a depth-wise 1D convolution layer. We show that the SVDF can be expressed as a special case of the S1DCNN layer. Experimental results show that the S1DCNN achieve 19.0\% relative false reject ratio (FRR) reduction with a similar model size and a similar time delay compared to the SVDF. By using longer time delays, the S1DCNN further improve the FRR up to 12.2\% relative.
\end{abstract}

\section{Introduction}
Speech is increasingly becoming a natural way to interact with consumer electronic devices. For privacy reasons, these device rely on the users to preface their commands with a target phrase prior to sending the continuous speech to the cloud for further understanding. Therefore, accurate on-device voice trigger detection is crucial to usability of these systems. Voice trigger detection has to run in an always-on fashion and in many cases on battery-powered devices such as phones or wearables such as earphones. Memory and compute efficiency play a key role in minimizing power consumption.

In recent years, with the advent of deep learning, neural networks have been used for voice trigger detection extensively. Earlier works ~\cite{panchapagesan2016multi,sun2017compressed,kumatani2017direct,guo2018time,wu2018monophone,prabhavalkar2015automatic,gruenstein2017cascade,team2017hey,tucker2016model,sigtia2018} used a hybrid approach where a deep neural network (DNN) was used to to estimate the observation probabilities over hidden Markov model (HMM) states. The HMM is used to compute the score for the correct sequence of states in the target phrase given the acoustic evidence. %Instead of the DNN acoustic model, a convolutional neural network was also used for KWS \cite{sainath2015convolutional}.
 More recently, however, end-to-end approaches have been used for voice trigger detection, where all components of the detection system are jointly optimized to directly produce the detection likelihood score. This end-to-end approach is in contrast to a suboptimal approach of optimizing independent components separately as used in the DNN-HMM systems. Various model architectures, including fully-connected networks \cite{chen2014small}, convolutional neural networks (CNNs), and recurrent neural networks ~\cite{fernandez2007application,he2017streaming,arik2017convolutional} \cite{sainath2015convolutional} have been explored for the end-to-end approach. While achieving improvements over the DNN-HMM based approaches, these newer techniques are computationally complex for embedded applications, where the voice trigger detection system should work on an always-on processor (AOP) with limited memory and power consumption.

To perform on-device voice trigger detection more efficiently, a low-rank approximation of a fully-connected layer was proposed in \cite{43813}. The approximation, called a singular value decomposition filter (SVDF), decomposes a weight matrix into two filters applied in feature and time dimensions, respectively. This decomposition reduces the number of model parameters and enables lightweight streaming voice trigger detection. The SVDF was then applied to end-to-end streaming voice trigger detection in \cite{alvarez2019end}, where the SVDF layers were stacked to enable a long receptive field and to directly estimate the triggering score. This approach enabled end-to-end streaming voice trigger detection with a small model size.

In this work, we propose an alternative approach for end-to-end voice trigger detection. We use stacked 1D CNN (S1DCNN) layers, where the first CNN layer combines information over the feature dimension, and the second CNN layer performs a depth-wise convolution in the time dimension. Similarly to the SVDF, the S1DCNN efficiently aggregates information both in the feature and the time dimensions. The S1DCNN can be implemented using CNN layers, and the length of the left/right receptive field can be  controlled via hyperparameter setting for the CNN layers. In addition, we show that the S1DCNN includes the SVDF as a special case with certain settings. The S1DCNN layers are stacked to have a reasonably long receptive field and perform end-to-end voice trigger detection.

%we believe optimizing independent components of system is suboptimal and train a novel efficient architecture end-to-end to detect presence of a keyword in continuous speech. Our evaluations demonstrate up to xx$\%$ improvement in accuracy with similar compute and memory cost.

Experimental evaluation on a voice trigger detection task shows that the S1DCNN outperforms the SVDF by 19.0\% relative in terms of the false reject ratio (FRR) with a similar model size. By exploring model parameter settings, the performance of the S1DCNN can further be improved by 12.2\% relative by introducing an additional time delay.

The rest of the paper is organized as follows. In Section \ref{sec:SVDF}, we review the singular value decomposition filter. Next, in Section \ref{sec:S1DCNN}, we describe the architecture of the proposed S1DCNN. Section \ref{sec:diff} compares the two approaches. Section \ref{sec:KWS} describes how we perform end-to-end voice trigger detection by using the S1DCNN.  In Section \ref{sec:exp}, we present the experimental setup, and the results of our evaluations. Finally, we conclude with a discussion of our findings in Section \ref{sec:conc}.

\section{Singular value decomposition filter}
\label{sec:SVDF}
Motivated by the structure in the filters learned at individual nodes in the first hidden layer of fully-connected neural networks, the authors of ~\cite{43813} proposed a low-rank approximation of the filter using singular value decomposition to reduce the number of parameters in the model. The first layer, which overlays its weights on input time-frequency representation, is replaced with a singular value decomposition filter (SVDF).

Let $x_{f,t} (t=1,...,T, f=1,..,F)$ denote an $f$-th value of an $F$-dimensional feature vector at time frame $t$. By concatenating the input vectors from $K$ frames, an input vector at each time frame for a standard fully-connected layer can be obtained as an $(F \times K)$-dimensional vector. Instead of combining information of the $(F \times K)$-dimensional vector at the same time with an $(F \times K)$-dimensional filter, the SVDF decomposes the process into two filtering processes in the feature and time dimensions, respectively.  The output activation, $a_t$, for each node in the SVDF layer at a given time frame $t$ is computed as:

\begin{equation}
a_t = g(\sum_{k=1}^{K}\alpha_k \sum_{f=1}^F\beta_f x_{f,(t-K+k)}),\label{eq:SVDF}
%a_t = g(\sum_{i=0}^{T-1}{\alpha_i}\sum_{j=1}^{F}{\beta_j}x_{(t−T+i),j})
\end{equation}
where $\alpha$ and $\beta$ denote filters of the SVDF node, and $g(\cdot)$ denotes an activation function. $\beta$ slides over input features with a stride of F, thus combining information in the feature vector into a single scalar. $\alpha$ mixes the resulting scalars for $K$ time steps into a single output value. This approach reduces the memory and compute complexity of the original layer from $O(F \times K)$ to $O(F + K)$ since an $(F \times K)$-dimensional filter is decomposed into an $F$-dimensional filter and a $K$-dimensional filter. In addition, this approach enables a stateful network that can memorize and utilize the past $K-1$ inference steps in making a decision at the current time frame.

In ~\cite{alvarez2019end}, SVDF layers are stacked to extend the receptive field of the network. Since there is no recurrent dependency, SVDF layers can operate on streaming input. If a network has $D$ stacked SVDF layers with memory size of $K$, it produces an output decision per incoming input frame while taking the past $D\times(K -1)$ frames  into account. Therefore, the model takes a large input receptive field into account and has been shown to be effective for end-of-keyword detection. Additionally, since SVDF layers are obtained through low-rank decomposition, the resulting network is highly compressed for deployment in resource-constrained settings.

Although the SVDF was proposed as a decomposition of a fully-connected layer, the SVDF can also be regarded as a factorization of a 2D CNN layer since the input can be regarded as a 2D feature map. This fact suggests that the SVDF can be implemented by decomposing a 2D CNN layer, which is realized by stacking 1D CNN layers. In Section \ref{sec:S1DCNN}, we describe the S1DCNN as an alternative approach for efficient voice trigger detection, then we describe its relationship to the SVDF in Section \ref{sec:diff}.

\section{Stacked 1D CNN}
\label{sec:S1DCNN}

\begin{figure}[t]
  \centering
  \includegraphics[width=\linewidth]{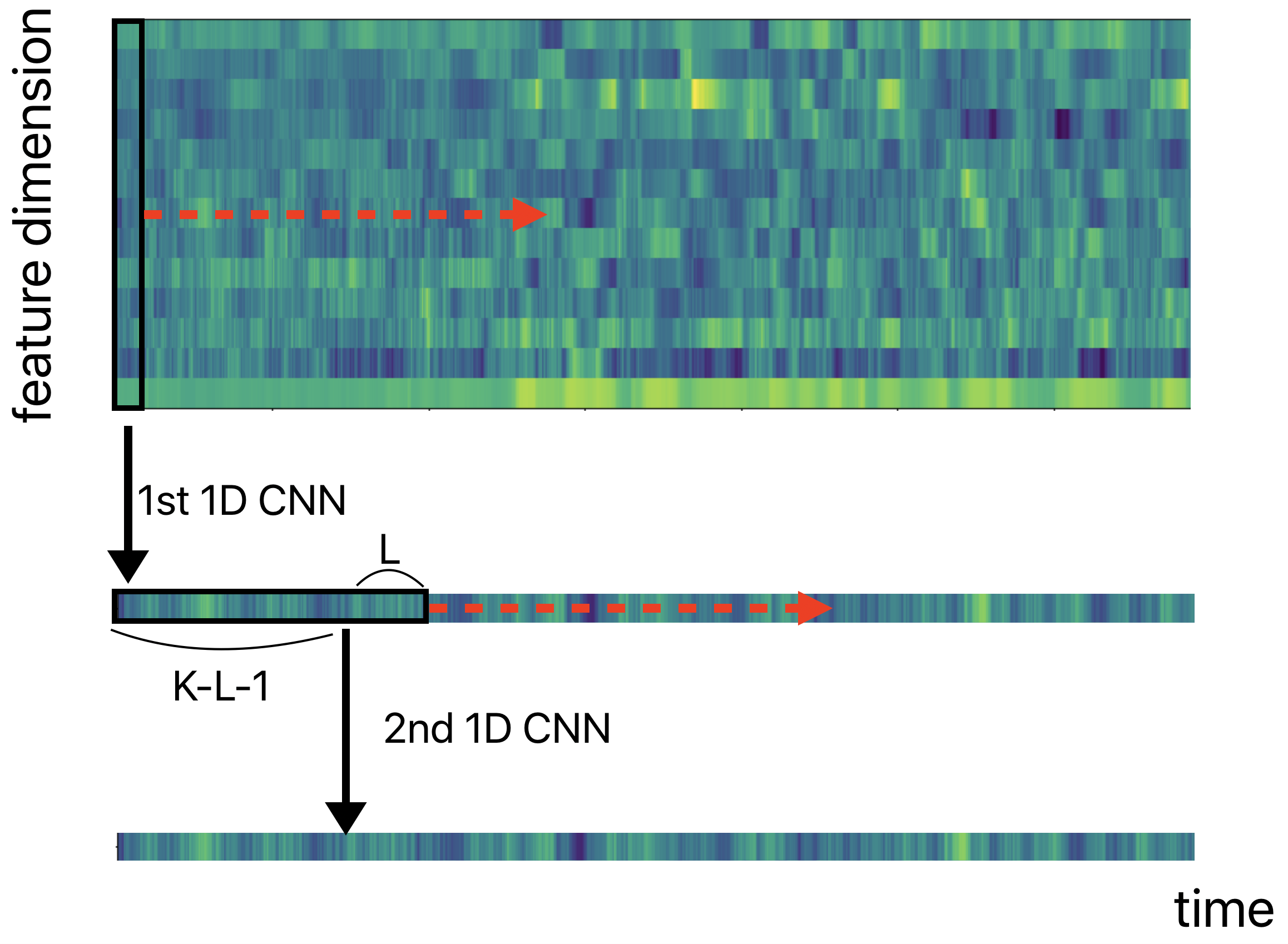}
  \caption{Computations by a stacked 1D CNN unit}
  \label{fig:S1DCNN}
\end{figure}

%We propose using the stacked 1D CNN (S1DCNN) as an alternative approach for end-to-end  voice trigger detection.
Figure \ref{fig:S1DCNN} illustrates  how an S1DCNN unit computes hidden outputs from inputs. The S1DCNN unit consists of a 1D convolution layer followed by a depth-wise 1D convolution layer.

Let us assume that the first 1D convolution layer has $N$ filters. Let $w_{f}^{(n)}$ denote the $n$-th CNN filter weight in dimension $f$ of an input vector $x_{1:T}$, and $b^{(n)}$ denote a bias parameter of the $n$-th filter, where $n \in (1, ..., N)$. The output of the $n$-th filter of the first 1D convolution layer can be written as
\begin{equation}
a_t^{(n)} = g^{(1st)}(\sum_{f=1}^{F}w_{f}^{(n)}x_{f,t} + b^{(n)}),\label{eq:first_CNN}
\end{equation}
where $g^{(1st)}(\cdot)$ denotes an activation function for the first 1D CNN layer. Thus, this layer accumulates information over the feature dimension.

Let us assume that the second 1D convolution layer has $N$ filters of size $K$. The second 1D convolution filter is applied to the outputs of the first 1D convolution layer in a so-called ``depth-wise'' manner, where the $n$-th filter of the second 1DCNN is  applied only to the $n$-th filter output of the first 1DCNN.
%Let us assume that the second 1D convolution layer has $N$ filters of size $K$. It means that the $n$-th filter of the second 1DCNN is applied only to the $n$-th filter output of the first 1DCNN.
 Let $w^{'}_{k}{}^{(n)}$ denote $k$-th component of $n$-th CNN filter weight, and $b^{'}{}^{(n)}$ denote a bias parameter of the $n$-th filter. By performing depth-wise 1D convolution, an output of the $n$-th filter can be written as
\begin{equation}
a^{'}_t{}^{(n)} = g^{(2nd)}(\sum_{k=1}^{K}w^{'}_{k}{}^{(n)}a_{t-K+k+L}^{(n)} + b^{'}{}^{(n)}),\label{eq:second_CNN}
\end{equation}
where $g^{(2nd)}(\cdot)$ denotes an activation function for the second 1D CNN layer, and $L$ denotes time offset. The second 1DCNN layer looks up $K-L-1$ left outputs as well as the current output and $L$ right outputs from the first layer. Therefore this layer accumulates information across the time frames and gives $N$-dimensional output vectors of size $T$. The length of future context can be controlled by $L$, where increasing $L$ introduces longer time delays in a streaming scenario.

Compared to a standard 2D CNN filter, the S1DCNN can be regarded as a factorization of a 2D CNN filter. An $F$ $\times$ $K$ filter of the 2D CNN layer is factorized into an $F$ $\times$ $1$ filter of the first 1D CNN layer and a $1$ $\times$ $K$ filter of the second 1D CNN layer. This factorization reduces the number of parameters from $\mathcal{O}(F \times K)$ to $\mathcal{O}(F + K)$. Since the SVDF can also be described as a decomposition of the 2D CNN layer, the SVDF and the S1DCNN have a tight relationship.

\section{Relationship between SVDF and S1DCNN}
\label{sec:diff}
With certain settings, the S1DCNN can be equivalent to the SVDF. Let us assume that $g^{(1st)}$ is an identity function:
\begin{equation}
x = g^{(1st)}(x).
\label{eq:f_first}
\end{equation}
From Eqs. (\ref{eq:first_CNN}), (\ref{eq:second_CNN}) and (\ref{eq:f_first}), the output of the S1DCNN can be rewritten as
\begin{equation}
a^{'}_t{}^{(n)} = g^{(2nd)}(\sum_{k=1}^{K}w^{'}_{k}{}^{(n)}(\sum_{f=1}^{F}w_{f}^{(n)}x_{f,t-K+k+L} + b^{(n)}) + b^{'}{}^{(n)}).\label{eq:S1DCNN}
\end{equation}
By setting $b^{(n)}$, $b^{'}{}^{(n)}$ and $L$ at $0$\footnote{In \cite{43813}, the SVDF was originally proposed with $L \geq 0$, however, it was used with $L=0$ in \cite{alvarez2019end} for end-to-end voice trigger detection. This paper reformulates the SVDF by using S1DCNN with $L \geq 0$, and effects of $L$ will be investigated in end-to-end voice trigger detection experiments. }, eq. (\ref{eq:S1DCNN}) is equivalent to eq. (\ref{eq:SVDF}) when $g^{(2nd)} = g$
\begin{equation}
a^{'}_t{}^{(n)} = g^{(2nd)}(\sum_{k=1}^{K}w^{'}_{k}{}^{(n)}\sum_{f=1}^{F}w_{f}^{(n)}x_{f,t-K+k}),\label{eq:S1DCNN2}
\end{equation}
where $\alpha_k $ and $\beta_f$ in the SVDF correspond to $w^{'}_{k}{}$ and $w_{f}$, respectively. This shows that the SVDF can be easily implemented as the S1DCNN, and the model performance can be potentially improved via hyperparameter settings over the bias parameters and the length of future context $L$.

\section{End-to-end voice trigger detection\\ based on S1DCNN}
\label{sec:KWS}
As in \cite{alvarez2019end}, the S1DCNN can also be stacked for end-to-end voice trigger detection. Let us assume that $C$ left and right frames are concatenated with the current frame to construct the feature vector of size $F \times (2C+1)$ for the current time step\footnote{Our preliminary experiments showed that frame concatenation was needed, even for the end-to-end models, to obtain sufficient performance. }. $D$ stacked S1DCNN layers are used for end-to-end voice trigger detection, where the model takes into account $(K-1-L) \times D$ past and $L \times D$ future frames of  features. Since each feature vector at a time step contains $C$ frames of context, the receptive field of the model is $(K-1-L) \times D+C$ past and $L \times D + C$ future frames. If $K$ and $D$ are sufficiently large, the  receptive field can cover an entire target phrase, which enables streaming voice trigger detection with a binary classifier.
% The target label for the end-to-end model can be obtained, e.g.,  by setting label 1 at the end of the target phrase, and label 0 at other time frames.
 The positive class label can be repeated across frames to avoid highly imbalanced target label distributions \cite{alvarez2019end}.

\section{Experimental evaluation}
\label{sec:exp}

\begin{table*}[t]
  \caption{False reject ratios at 1 false alarm per hour}
  \label{tab:FRRs}
  \centering
%  \resizebox{\columnwidth}{
  \begin{tabular}{ r lr |r |r |r |r |r |r }
    \toprule
    \multicolumn{1}{c}{\textbf{Models}} &
%                                         \multicolumn{1}{c}{\textbf{$g^{(1st)}$}} &
 %                                        \multicolumn{1}{c}{\textbf{$g^{(2nd)}$}} &
                                          \multicolumn{1}{c}{\textbf{E2E}} &
                                         \multicolumn{1}{c}{\textbf{bias param.}} &
                                         \multicolumn{1}{c}{\textbf{L}} &
                                         \multicolumn{1}{c}{\textbf{recpt. field (left/right) [ms]}} &
                                         \multicolumn{1}{c}{\textbf{\# param.}} &
                                          \multicolumn{1}{c}{\textbf{\# MACs}} &
                                         \multicolumn{1}{c}{\textbf{FRRs [\%]}}\\
    \midrule
    DNN-HMM \cite{sigtia2018} &&  &&&13979&$\sim$12.8k&4.99~~~             \\
    SVDF \cite{alvarez2019end}                   &\checkmark    &&0&610 / 50&13993&$\sim$13.0k&3.95~~~               \\
    \midrule
    S1DCNN                    &  & &0&610 / 50&&&3.20~~~       \\
  && &1&540 / 120&&&\textbf{2.81}~~~       \\
  &\checkmark  &\checkmark &2&470 / 190&14441&$\sim$13.0k&3.15~~~       \\
 & & &3&400 / 260&&&3.77~~~       \\
   & & &4&330 / 330&&&4.33~~~       \\
    \bottomrule
  \end{tabular}
 % }
  \label{tab:FRRs}
\end{table*}

We evaluated the performance of the S1DCNN on a voice trigger detection task by comparing with a DNN-HMM hybrid system and the SVDF end-to-end model. We also investigated the effect of the length of future context used in the S1DCNN.

\subsection{Data}
For training, we used $\sim$500k utterances in English recorded anonymously with smart phones and tablet devices. Each utterance starts with a specific target phrase, i.e., ``Hey Siri'',  followed by a query to the voice assistant. The utterances were augmented by convolving room impulse responses and by adding echo residuals. Next, gain augmentation was performed by changing the gain by $P$ dB, where different $P$ was randomly chosen from $-32$, $-20$ and $0$ for different utterances. This gain augmentation was performed for both closing the gap in audio volumes between server-side (training) and on-device (inference) data, and making models more robust to various volumes. For end-to-end model training, we randomly drop the target phrase portion from the audio with a 50\% chance to create utterances that do not contain the target phrase (negative samples).

For evaluation, we used 6509 utterances containing the target phrase as positive samples. As negative samples, we used $\sim$2700 hours of audio in a range of noise conditions without the trigger phrase.

\subsection{Settings}

We used 13-dimensional mel-frequency cepstral coefficients (MFCCs) as input features. The MFCCs were extracted with 25ms window and 10ms shift. For the baseline DNN-HMM system, features at 9 past frames and 9 future frames were concatenated with the current frame, which resulted in 247-dimensional input vector at each time frame. For the end-to-end models, we concatenated 5 past frames and 5 future frames with the current frame. We used fewer context frames for the end-to-end models because the stacked CNNs extend the receptive field as described in Section \ref{sec:KWS} and the larger input dimension increases the number of parameters of the first layer unnecessarily.

We used two baseline models for comparison. One was a DNN-HMM hybrid system and the other was an SVDF end-to-end model \cite{alvarez2019end}. The DNN consisted of 5 fully-connected layers with 32 units followed by a linear layer, where a batch normalization layer was applied after each fully-connected layer. The sigmoid activation function was applied after each batch normalization layer. The last linear layer transformed 32-dimensional hidden outputs into 20-dimensional logits for 20 target classes, then softmax was applied to obtain probabilities for the classes from the logits. The 20 target classes consisted of a silence class, a general speech class, and 18 classes for the target phrase. The 18 classes for the target phrase were defined with 6 tri-phones, where each tri-phone had 3 states (classes). With these settings, the DNN had 13979 parameters. See \cite{sigtia2018} for more details of the baseline DNN-HMM hybrid voice trigger detection system.

The SVDF model consisted of 7 SVDF layers, each of which had 32 filters. The rectified linear unit (ReLU) activation was applied as $g(\cdot)$ in each SVDF layer. A batch normalization layer was applied after each SVDF layer. The memory size $K$ was set to $9$. A linear layer was used to transform 32-dimensional hidden outputs into 2-dimensional logits for target and non-target classes. The softmax function converted the logits into probabilities for these two classes.  We set labels at 30 frames before the end time of the target phrase as the target class, and the rest as the non-target class,  similar to \cite{alvarez2019end}. During inference, we computed the average of outputs at 29 past frames and the current frame to get the final score.

The proposed S1DCNN had the same model architecture as the SVDF model except for the differences described in section \ref{sec:diff}. As a result, the receptive field of the model was $((8-L) \times 7 + 5) \times 10$ ms before and $(L \times 7 + 5) \times 10$ ms after the current time frame. We used the identity and the ReLU functions for $g^{(1st)}(\cdot)$ and $g^{(2nd)}(\cdot)$, respectively. The same binary labels were used for training, and the output averaging was also applied during inference. With these settings, the SVDF and the S1DCNN had similar model sizes and the number of multiplier–accumulator (MAC) operations compared to the baseline DNN (see table \ref{tab:FRRs}).

We varied the length of the future context  $L$ for the S1DCNN from 0 to 4 to investigate its effect on the voice trigger detection performance. With these settings,  the S1DCNN had an $(L \times 7 + 5) \times 10$ ms time delay.
%  Moreover, an effect of a model size was also investigated by setting the numbers of filters at 22 and 26 for both the SVDF and the S1DCNN in the case of $L=0$.

Adam optimizer \cite{kingma2014adam} was used for model training. The initial learning rate was set at 0.001, betas were set at [0.9,0.999]. We used a minibatch size of 256 for training. Our training consisted of warm-up and main stages. In the warm-up stage, the learning rate was increased by a factor of 1.4 when the cross validation loss decreased at the end of each epoch. Once the cross validation loss no longer decreased for 8 consecutive epochs, we rolled back to the best performing model, and moved onto the main stage. In the main stage, a learning rate decay of 0.5 was applied when the cross validation loss did not decrease for 4 consecutive epochs. Early-stopping of training was applied when we did not see cross validation loss decrease for 8 consecutive epochs.  The size of an epoch for the warm-up and the main stages were 100k and 500k utterances, respectively. All models were 32 bit floating point format, and performance evaluation of quantized models will be our future work.

\subsection{Results}
%\subsubsection{Model comparison}

\begin{figure}[t]
  \centering
  \includegraphics[width=\linewidth]{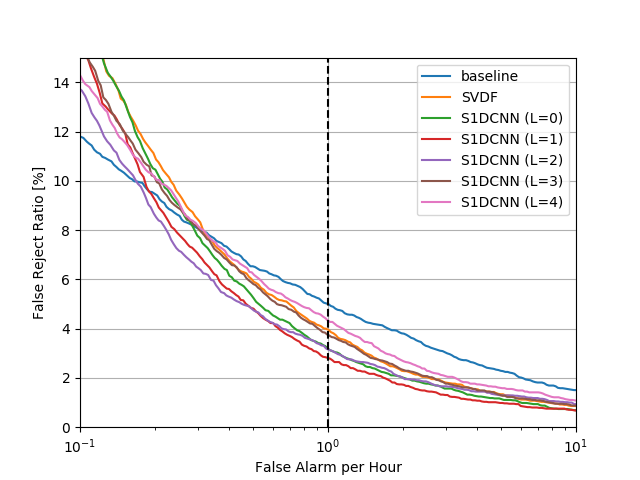}
  \caption{DET curves obtained by the baseline DNN, SVDF, and S1DCNN. The vertical dot line indicates the operating point.}
  \label{fig:DET}
\end{figure}

Figure \ref{fig:DET} shows detection error tradeoff (DET) curves obtained by the DNN-HMM, the SVDF and the S1DCNNs. Table \ref{tab:FRRs} shows FRRs at an operating point, i.e., 1 false alarm per hour. Compared with the DNN-HMM, the SVDF end-to-end model achieved 20.8\% relative FRR reduction. By simply adding the bias parameters, the S1DCNN with $L=0$ achieved a further performance gain by 19.0\% relative to the SVDF. By allowing an additional time delay, the FRR was further improved by 12.2\% relative to the S1DCNN with $L=0$, although the best model with $L=1$ had $(1 \times 7) \times 10$ ms additional time delay compared to the SVDF and the S1DCNN with $L=0$. These results show that the S1DCNN outperforms the SVDF with a similar model size, and can further be improved by allowing an additional time delay. The reason for worse performances obtained with $L \geq 3$ would be too short left receptive fields (330/400 ms) to capture an entire target phrase. Note that it is feasible to increase the right context length while keeping the sufficient left context length, although that results in increasing the model size.

%\subsubsection{Effect by model size}

%\begin{figure}[t]
%\setlength{\abovecaptionskip}{5pt}
 % \centering
  %\includegraphics[width=\linewidth]{Sizes_v2.png}
  %\caption{FRRs at 1 false alarm per hour obtained with changing model sizes.}
  %\label{fig:size}
%\end{figure}

%Figure \ref{fig:size} shows the FRRs at 1 false alarm per hour obtained by changing model sizes of the SVDF and the S1DCNN with $L=0$. Interestingly, the SVDF kept its performance even when its model size being reduced by 40\% relative. On the other hand, the performance by the S1DCNN degraded with the reducing model sizes as expected. Keeping the performance by the S1DCNN even with smaller model sizes will be one of our future work.

\section{Conclusions}
\label{sec:conc}
In this paper, we propose the stacked 1D convolutional neural network (S1DCNN) as an alternative approach for end-to-end voice trigger detection. The S1DCNN layer consists of a 1D convolution layer followed by a depth-wise 1D convolution layer, which can efficiently perform voice trigger detection with the small number of model parameters from streaming audio signals. The S1DCNN includes the previously-proposed SVDF as an special case. Experimental results showed that the S1DCNN outperformed the SVDF by 19.0\% relative with a similar model size and a time delay. By allowing an additional time delay, the performance of the S1DCNN further improved by up to 12.2\% relative.
% Our future work includes keeping a KWS performance by the S1DCNN with even smaller model sizes.

%\section{Acknowledgements}

\bibliographystyle{IEEEtran}

\bibliography{mybib}

% \begin{thebibliography}{9}
% \bibitem[1]{Davis80-COP}
%   S.\ B.\ Davis and P.\ Mermelstein,
%   ``Comparison of parametric representation for monosyllabic word recognition in continuously spoken sentences,''
%   \textit{IEEE Transactions on Acoustics, Speech and Signal Processing}, vol.~28, no.~4, pp.~357--366, 1980.
% \bibitem[2]{Rabiner89-ATO}
%   L.\ R.\ Rabiner,
%   ``A tutorial on hidden Markov models and selected applications in speech recognition,''
%   \textit{Proceedings of the IEEE}, vol.~77, no.~2, pp.~257-286, 1989.
% \bibitem[3]{Hastie09-TEO}
%   T.\ Hastie, R.\ Tibshirani, and J.\ Friedman,
%   \textit{The Elements of Statistical Learning -- Data Mining, Inference, and Prediction}.
%   New York: Springer, 2009.
% \bibitem[4]{YourName17-XXX}
%   F.\ Lastname1, F.\ Lastname2, and F.\ Lastname3,
%   ``Title of your INTERSPEECH 2020 publication,''
%   in \textit{Interspeech 2020 -- 20\textsuperscript{th} Annual Conference of the International Speech Communication Association, September 15-19, Graz, Austria, Proceedings, Proceedings}, 2020, pp.~100--104.
% \end{thebibliography}

\end{document}